\documentclass[12pt]{iopart}
\usepackage[dvips]{graphicx}
\begin{document}

%\linespread{1.6}

\title[Density fluctuations on mm and Mpc scales]
{A comparative study of density fluctuations on mm and Mpc scales}

\author{N P Basse \footnote{Current address: ABB Corporate Research, Segelhofstrasse 1, CH-5405 Baden-D\"attwil, Switzerland}}

\address{Plasma Science and Fusion Center, Massachusetts Institute of Technology, Cambridge, MA 02139, USA}
\eads{\mailto{basse@psfc.mit.edu}}

\begin{abstract}
We have in earlier work (Basse N P 2005 {\it Phys. Lett. A} {\bf
340} 456) reported on intriguing similarities between density
fluctuation power versus wavenumber on small (mm) and large (Mpc)
scales.

In this paper we expand upon our previous studies of small and
large scale measurements made in fusion plasmas and using
cosmological data, respectively.

The measurements are compared to predictions from classical fluid
turbulence theory. Both small and large scale data can be fitted
to a functional form that is consistent with the dissipation range
of turbulence.

The comparable dependency of density fluctuation power on
wavenumber in fusion plasmas and cosmology might indicate a common
origin of these fluctuations.
\end{abstract}

\submitto{\PPCF}
\maketitle

\section{Introduction}
\label{sec:intro}

Transport of particles and energy across the confining magnetic
field of fusion devices is anomalous \cite{wootton}, i.e., it is
much larger than the neoclassical transport level associated with
binary collisions in a toroidal geometry \cite{hinton}. It is
thought that anomalous transport is caused by plasma turbulence,
which in turn manifests itself as fluctuations in most plasma
parameters. To understand anomalous transport, a two-pronged
approach is being applied: (i) sophisticated diagnostics measure
fluctuations and (ii) advanced simulations are being developed and
compared to these measurements. Once our understanding of the
relationship between fluctuation-induced anomalous transport and
plasma confinement quality is more complete, we will be able to
reduce transport due to the identified mechanism(s).

The fusion plasma measurements presented in this paper are of
fluctuations in the electron density. Small-angle collective
scattering \cite{saffman,basse1} was used in the Wendelstein 7-AS
(W7-AS) stellarator \cite{renner} and phase-contrast imaging (PCI)
\cite{mazurenko} is being used in the Alcator C-Mod tokamak
\cite{hutch}.

We specifically study density fluctuation power versus wavenumber
(also known as the wavenumber spectrum) in W7-AS and C-Mod. These
wavenumber spectra characterize the nonlinear interaction between
turbulent modes having different length scales.

The second part of our measurements, wavenumber spectra (i) of
galaxies from the Sloan Digital Sky Survey (SDSS) \cite{sdss} and
(ii) from a variety of sources (including the SDSS data) are
published in Ref. \cite{tegmark1} and have been made available to
us \cite{tegmark2}.

The paper is organized as follows: In section \ref{sec:rev} we
review our initial results from Ref. \cite{basse2}. Thereafter we
analyze our expanded data set in section \ref{sec:add} and in
response to the results revise our treatment of the original W7-AS
measurements in section \ref{sec:anal}. A discussion follows in
section \ref{sec:disc} and we conclude in section \ref{sec:conc}.

\section{Review of earlier findings}
\label{sec:rev}

\subsection{Stellarator fusion plasmas} \label{subsec:fusion}

In Ref. \cite{basse2} we proposed that the density fluctuation
power $P$ decreases exponentially with increasing wavenumber $k$
on mm scales in fusion plasmas according to

\begin{equation}
P(k) \propto \frac{1}{k} \times e^{-nk}, \label{eq:exp_decay}
\end{equation}

\noindent where $n > 0$ is a constant having a dimension of length
and $k = 2\pi/\lambda$, where $\lambda$ is the corresponding
wavelength. Initially we suggested the simplified form

\begin{equation}
P(k) \propto e^{-nk} \label{eq:exp_decay_simpl}
\end{equation}

\noindent in Ref. \cite{basse3}. Eq. (\ref{eq:exp_decay_simpl})
also holds for density fluctuations in the Tore Supra tokamak
\cite{hennequin}.

\begin{center}
\begin{figure}
\includegraphics[width=12cm]{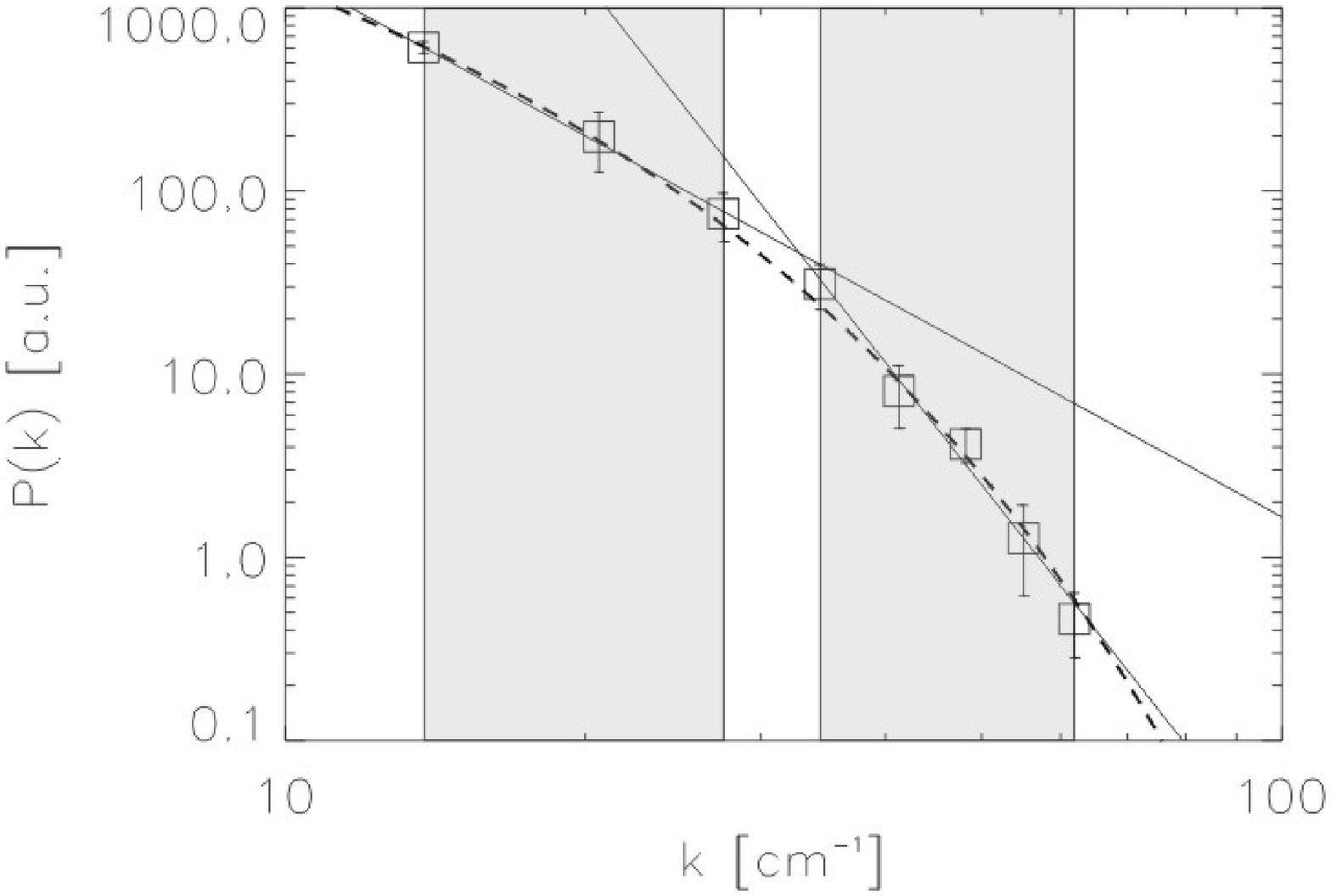}
\caption{\label{fig:w7as} Wavenumber spectrum of broadband
turbulence in W7-AS. Squares are measured points. Solid lines are
power-law fits (Eq. (\ref{eq:pow_decay})) and the dashed line is a
fit to Eq. (\ref{eq:exp_decay}). The power-law fit grouping is
indicated by the semi-transparent rectangles, the fit to Eq.
(\ref{eq:exp_decay}) uses all data points.}
\end{figure}
\end{center}

A wavenumber spectrum of turbulence in W7-AS is shown in Fig.
\ref{fig:w7as}. The measured points are shown along with two
power-law fits

\begin{equation}
P(k) \propto k^{-m}, \label{eq:pow_decay}
\end{equation}

\noindent where $m$ is a dimensionless constant. The power-law
fits are shown as solid lines and an exponential fit to Eq.
(\ref{eq:exp_decay}) is shown as a dashed line.

All fits shown in this paper have a normalized $\chi^2$ $\le$ 1,
ensuring the quality of the fits is good. The error bars are
standard deviations and the semi-transparent rectangles indicate
which points are included to make the fits.

The power-law fits are motivated by classical fluid turbulence
theory where one expects wavenumber spectra to exhibit power-law
behavior with exponents $m$ depending on the dimension of the
observed turbulence:

\begin{itemize}
\item 3D: Energy is injected at a large scale and redistributed
(cascaded) by nonlinear interactions down to a small dissipative
scale. In this case, the energy spectrum in the inertial range
$E(k) \propto k^{-5/3}$ \cite{frisch}. \item 2D: Here, two
power-laws exist on either side of the energy injection scale. For
smaller wavenumbers, the inverse energy cascade obeys $E(k)
\propto k^{-5/3}$ and for larger wavenumbers, the enstrophy
cascade follows $E(k) \propto k^{-3}$ \cite{antar1}. \item 1D:
Energy is injected at a large scale and dissipated at a small
scale; $E(k) \propto k^{-2}$ \cite{neumann}.
\end{itemize}

Our measured power is equivalent to the $d$-dimensional energy
spectrum $F_d(k)$ \cite{tennekes,frisch,antar2}

\begin{eqnarray}
P(k) = F_d(k) = \frac{E(k)}{A_d} \nonumber \\ \nonumber \\ A_1 = 2
\hspace{2cm} A_2 = 2\pi k \hspace{2cm} A_3 = 4\pi k^2,
\label{eq:e_spec}
\end{eqnarray}

\noindent where $A_d$ is the surface area of a sphere having
radius $k$ and dimension $d$. Usually on would assume that $d = 2$
in fusion plasmas, since transport along magnetic field lines is
nearly instantaneous. The fits to Eq. (\ref{eq:pow_decay}) in Fig.
\ref{fig:w7as} yield exponents $m$ = 3.0 $\pm$ 0.4 (small
wavenumbers) and 6.9 $\pm$ 0.7 (large wavenumbers). A similar
behavior has previously been reported in Ref. \cite{honore} where
it was speculated that the wavenumber value at the transition
between the two power-laws should correspond to a characteristic
spatial scale in the plasma. The only length scale close to the
transitional value was found to be the ion Larmor radius $\rho_i$.

The spectrum at small wavenumbers is roughly consistent with the
inverse energy cascade in 2D turbulence, $F_2(k) \propto
k^{-8/3}$. The exponent at large wavenumbers does not fit into
this framework. However, for very large wavenumbers one enters the
dissipation range; here, it has been argued that the energy
spectrum could have one of the following dependencies

\begin{equation}
E_{\rm Neumann}(k) \propto e^{-ak} \hspace{2cm} E_{\rm
Heisenberg}(k) \propto k^{-7}, \label{eq:diss}
\end{equation}

\noindent where $a > 0$ is a constant having a dimension of length
(see Ref. \cite{neumann} and references therein). The energy
spectrum proposed by J von Neumann was what initially inspired us
to investigate an exponential decay of $P(k)$ in Ref.
\cite{basse3}. J von Neumann's work on this topic is from 1949 and
two years later, A A Townsend published a more generalized
expression \cite{townsend}.

Fitting all wavenumbers to Eq. (\ref{eq:exp_decay}), $E_{\rm
Neumann}(k)/A_2$, we find that $n$ = 0.11 $\pm$ 0.004 cm or a
wavenumber of 57.1 cm$^{-1}$. Alternatively, the transitional
wavenumber found at the intersection between the two power-laws is
32.8 cm$^{-1}$ (0.19 cm). The expression $E_{\rm
Heisenberg}(k)/A_2$ yields $m$ = 8, which is close to the
experimental value $m$ = 6.9 for large wavenumbers. Calculating
the ion Larmor radius at the electron temperature $\rho_s$ for
this case we find that it is 0.1 cm, i.e. the same order of
magnitude as the spatial scales found above. We used $\rho_s$
instead of $\rho_i$ because ion temperature measurements were
unavailable.

Currently we can think of three possible explanations for the
behavior of the wavenumber spectrum:

\begin{enumerate}
\item We observe 2D turbulence and the transition between the two
power-laws occurs at a spatial scale where the inverse energy
cascade develops into the dissipation range. However, the
enstrophy cascade is not accounted for in this case. \item We
observe 2D turbulence in the dissipation range described by Eq.
(\ref{eq:exp_decay}) as proposed by J von Neumann.
\item Turbulence theory does not apply. The transition between two
power-laws or the characteristic scale found using a single
exponential function (Eq. (\ref{eq:exp_decay_simpl})) indicates
that one scale dominates the turbulent dynamics in the wavenumber
range studied.
\end{enumerate}

In measurements of 2D fluid turbulence, it has been demonstrated
that the inverse energy and forward enstrophy cascades merge into
a single power-law when the system transitions to being fully
turbulent \cite{shats}. This might be the reason for the missing
enstrophy cascade discussed in item (i) above.

\subsection{Cosmology} \label{subsec:galax}

The shape of the wavenumber spectrum shown in Fig. \ref{fig:w7as}
bears a striking resemblance to measurements of fluctuations in
the distribution of galaxies presented in Ref. \cite{tegmark1},
see Fig. \ref{fig:sdss}. This motivates us to apply the analysis
described in section \ref{subsec:fusion} to the galaxy data. In
section \ref{subsubsec:infla} we briefly put these measurements in
context and then present an analysis of the galaxy wavenumber
spectrum in section \ref{subsubsec:wano}.

\subsubsection{Inflation} \label{subsubsec:infla}

Dramatic developments have taken place in cosmology over the last
decade, lending increasing support to the paradigm of inflation as
an explanation for what took place before the events described by
the big bang theory \cite{guth}. Inflation solved the so-called
horizon and flatness problems, but was at odds with earlier
observations indicating that the ratio of the mass density of the
universe to the critical value, the density parameter $\Omega$,
was 0.2-0.3, while inflation predicted it should be 1:

\begin{eqnarray}
\Omega = \frac{\rho}{\rho_c} \nonumber \\ \nonumber \\ \Omega < 1:
{\rm open} \hspace{2cm} \Omega = 1: {\rm flat} \hspace{2cm} \Omega
> 1: {\rm closed},
\label{eq:mass}
\end{eqnarray}

\noindent where $\rho_c = 3H_0^2/8\pi G$ is the critical mass
density, $H_0$ $\simeq$ 70 km/s/Mpc is the Hubble parameter
observed today and $G$ is I Newton's gravitational constant
\cite{kinney}. However, new measurements in the late 1990's lead
to a drastic modification of $\Omega$: Observations of type Ia
supernovae (SN) showed that the separation velocity between
galaxies was speeding up, not slowing down as would be expected
for an open universe. The underlying explanation for this
accelerated expansion is not understood, but it seems that the
universe contains large quantities of negative pressure substance,
creating a gravitational repulsion driving the expansion. This
negative pressure material is called dark energy, the total
density of dark energy $\Omega_{\Lambda}$ is 0.7. The existence of
dark energy is equivalent to the cosmological constant $\Lambda$
introduced by A Einstein. The dark matter density $\Omega_d$ is
0.25 and the baryonic matter density $\Omega_b$ is 0.05, so the
total density is very close (or equal) to the critical density.
The SN Ia data is supported by measurements of nonuniformities in
the cosmic microwave background (CMB) radiation. The CMB
anisotropy is due to the presence of tiny primordial density
fluctuations at the time of recombination, where atoms formed. At
that point in time the age of the universe was about 400,000 years
and the temperature was 3000 K. The structures observed in the CMB
are called acoustic peaks, and the simplest versions of inflation
all reproduce these structures quite accurately. The acoustic
peaks can not be modelled by assuming that the universe is open.

\subsubsection{SDSS wavenumber spectrum} \label{subsubsec:wano}

\begin{center}
\begin{figure}
\includegraphics[width=12cm]{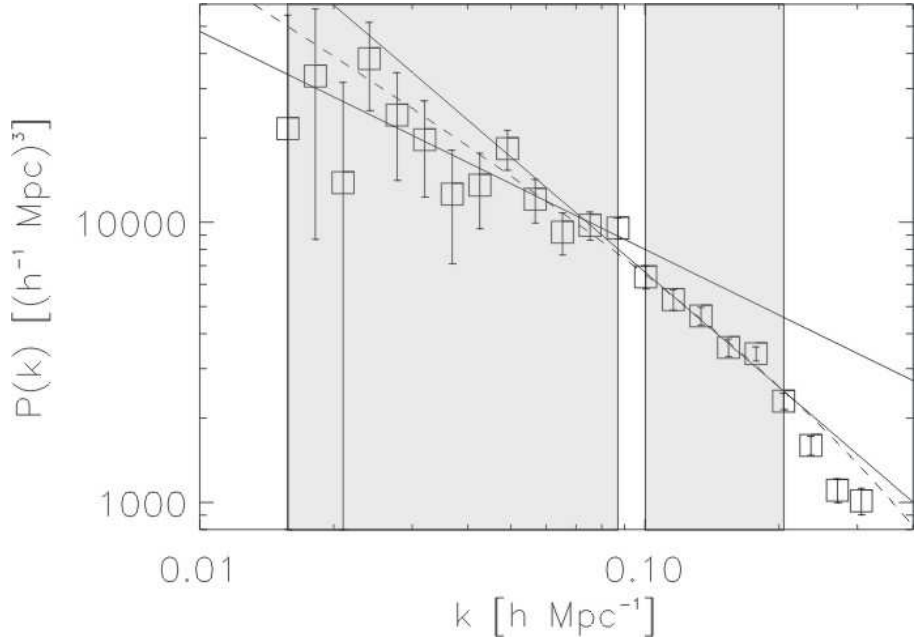}
\caption{\label{fig:sdss} Wavenumber spectrum of galaxies measured
by the SDSS Team. Squares are measured points. Solid lines are
power-law fits (Eq. (\ref{eq:pow_decay})), the dashed line is a
fit to Eq. (\ref{eq:exp_decay}). The power-law fit grouping of
points is chosen so the combined $\chi^2$ of the fits is
minimized. The power-law fit grouping is indicated by the
semi-transparent rectangles, the fit to Eq. (\ref{eq:exp_decay})
uses all data points up to $k$ = 0.2 h Mpc$^{-1}$. The data set is
taken from Ref. \cite{tegmark1}.}
\end{figure}
\end{center}

A study of density fluctuations on large scales using 205,443
galaxies has been published by the SDSS Team in Ref.
\cite{tegmark1}, see Fig. \ref{fig:sdss}. 3D maps of the universe
are provided by the SDSS galaxy redshift survey, observing about a
quarter of the celestial sphere using a 2.5 m telescope and a
charge-coupled device (CCD) camera. The galaxies had a mean
redshift $z \approx$ 0.1, corresponding to light emitted 1-2 Gyr
ago \cite{kinney}. Fixing some cosmological parameters to
Wilkinson Microwave Anisotropy Probe (WMAP) satellite values
\cite{spergel} one finds - using physics based models - that the
wavenumber spectrum measurements were fitted by a matter density
$\Omega_m = \Omega_d + \Omega_b = 0.295 \pm 0.0323$. In this case
$h = H_0$/(100 km/s/Mpc) = 0.72 was assumed and it was observed
that the wavenumber spectrum was not well characterized by a
single power-law.

A follow-up paper by the SDSS Team, Ref. \cite{tegmark3}, combined
non-CMB measurements (SDSS) with CMB measurements (WMAP) to
constrain free parameters of cosmological models and break CMB
degeneracies in parameter space. This resulted in $\Omega_m = 0.30
\pm 0.04$ and $h = 0.70_{-0.03}^{+0.04}$. Adding the SDSS
information more than halved WMAP-only error bars on some
parameters, e.g. the Hubble parameter and matter density.

The data presented in Fig. \ref{fig:sdss} has been taken from M
Tegmark's homepage \cite{tegmark4}. According to the
recommendation by the SDSS Team \cite{tegmark1}, the three largest
wavenumbers shown are not used in the fits described in this
section.

As we did for the W7-AS data in Section \ref{subsec:fusion}, we
fit the SDSS measurements to two power-laws (Eq.
(\ref{eq:pow_decay})) or the exponential function with an
algebraic pre-factor (Eq. (\ref{eq:exp_decay})). The power-law
fits are shown as solid lines, the exponential fit as a dashed
line.

The power-law fits yield exponents $m$ = 0.8 $\pm$ 0.03 (small
wavenumbers) and 1.4 $\pm$ 0.1 (large wavenumbers). The wavenumber
ranges were determined by minimizing the combined $\chi^2$ of the
fits. As the SDSS Team found, a single power-law can not describe
the observations. The exponents are not close to the ones
governing fluid turbulence discussed in section
\ref{subsec:fusion}. The transitional wavenumber (power-law
intersection) is 0.07 h Mpc$^{-1}$, corresponding to a length of
89.8 h$^{-1}$ Mpc.

We find the characteristic length from Eq. (\ref{eq:exp_decay}) to
be $n$ = 2.3 $\pm$ 0.5 h$^{-1}$ Mpc or a wavenumber of 2.7 h
Mpc$^{-1}$.

\section{Analysis of additional measurements}
\label{sec:add}

In Ref. \cite{basse2} we noted that both fusion plasma and
cosmological wavenumber spectra peak at small wavenumbers and
decrease both above and below that peak. In this section we
analyze supplemental measurements on a wider range of scales. The
fusion plasma measurements were made in a tokamak, so our explicit
assumption is that turbulence in stellarators and tokamaks is
comparable.

We fit the data to a modified version of Eq. (\ref{eq:exp_decay}),
namely

\begin{equation}
P(k) \propto \frac{1}{k^{\alpha}} \times e^{-nk},
\label{eq:exp_pow_decay}
\end{equation}

\noindent where $\alpha$ is brought in as an additional fit
parameter. The introduction of $\alpha$ is based on the assumed
functional form of the energy spectrum in the dissipation range of
fluid turbulence \cite{chen}. Further, since we have no clear
criteria for choosing between Eqs. (\ref{eq:exp_decay}) and
(\ref{eq:exp_decay_simpl}), leaving $\alpha$ to be fitted will
test our biased opinion in Ref. \cite{basse2} for Eq.
(\ref{eq:exp_decay}). Using Eq. (\ref{eq:e_spec}), Eq.
(\ref{eq:exp_pow_decay}) implies that $E(k)$ $\propto$
$k^{(d-1)-\alpha} \times e^{-nk}$.

\subsection{Tokamak fusion plasma}
\label{subsec:pci}

\begin{center}
\begin{figure}
\includegraphics[width=12cm]{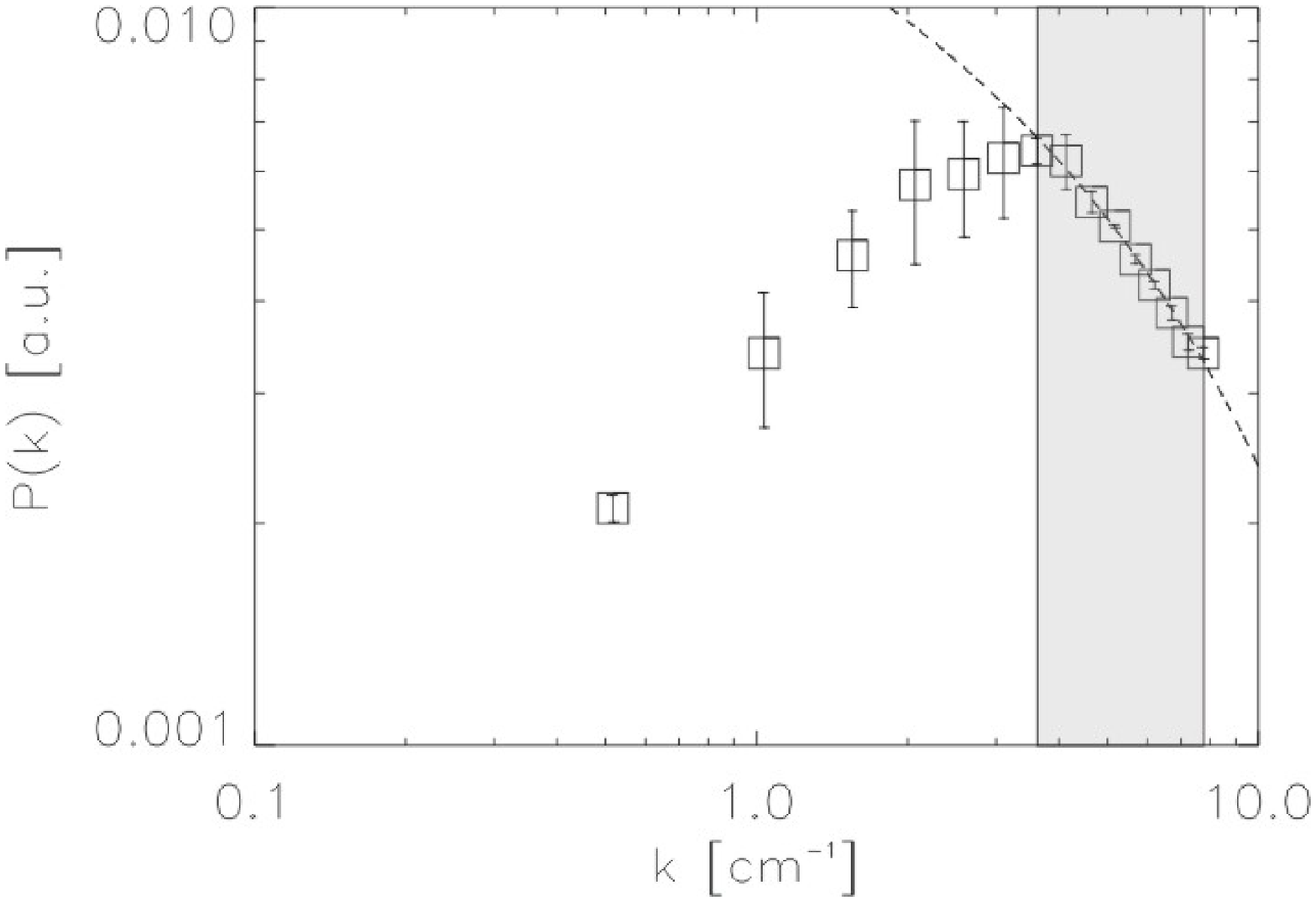}
\caption{\label{fig:pci} Wavenumber spectrum of broadband
turbulence in C-Mod. Squares are measured points. The dashed line
is a fit to Eq. (\ref{eq:exp_pow_decay}) using the measurements
indicated by the semi-transparent rectangle.}
\end{figure}
\end{center}

We analyze density fluctuations at small wavenumbers in C-Mod
using the PCI diagnostic. Measurements in the low confinement mode
taken from Fig. 11 in Ref. \cite{basse4} are shown in Fig.
\ref{fig:pci}. The dashed line shows the fit made using Eq.
(\ref{eq:exp_pow_decay}) which yields $n$ = 0.11 $\pm$ 0.005 cm
and $\alpha$ = 0.30 $\pm$ 0.02. The size of $n$ is close to the
value of $\rho_s$ in the given plasma.

\subsection{Cosmological wavenumber spectrum}
\label{subsec:gal_wano}

\begin{center}
\begin{figure}
\includegraphics[width=12cm]{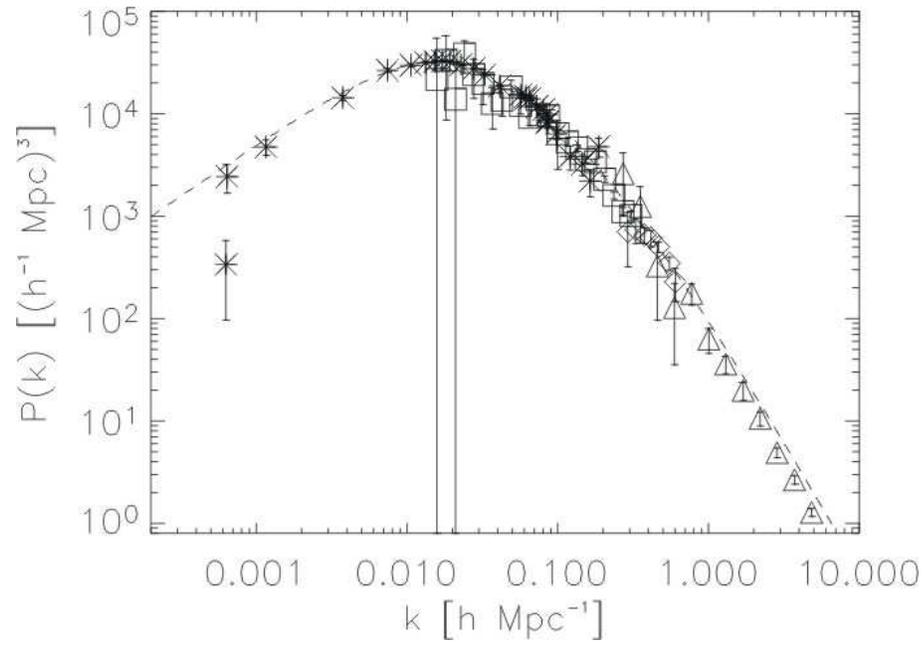}
\caption{\label{fig:cosmo_allk} Wavenumber spectrum of the
combined cosmological measurements. Asterisks are CMB radiation,
squares are SDSS, plus is cluster abundance, diamonds are weak
lensing and triangles are Lyman-$\alpha$ forest measurements. The
dashed line is a fit to the $\Lambda$CDM model. The data sets are
taken from Ref. \cite{tegmark1}.}
\end{figure}
\end{center}

The expanded cosmological data set shown in Fig. 38 of Ref.
\cite{tegmark1} is re-plotted in Fig. \ref{fig:cosmo_allk}. These
measurements have been provided by M Tegmark \cite{tegmark2}. They
are a combination of the measured density fluctuation power using
several diagnostic techniques; the dashed line is a fit to a
current model in cosmology, the so-called "vanilla light" flat
adiabatic $\Lambda$CDM model with negligible tilt, running tilt,
tensor modes or massive neutrinos \cite{tegmark1,tegmark3}.

\begin{center}
\begin{figure}
\includegraphics[width=12cm]{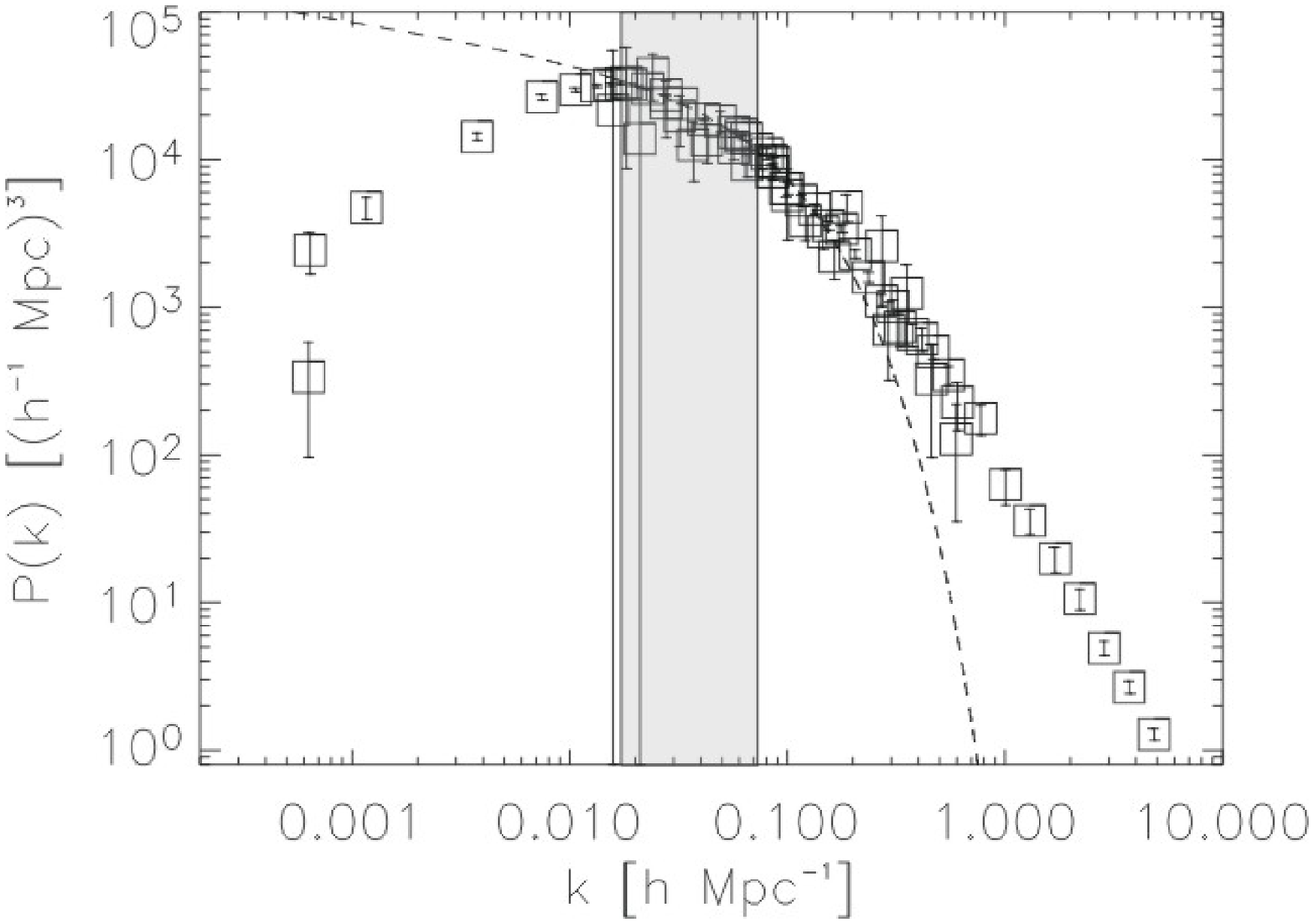}
\caption{\label{fig:cosmo_smallk} Wavenumber spectrum of the
combined cosmological measurements. Squares are measured points.
The dashed line is a fit to Eq. (\ref{eq:exp_pow_decay}) using the
data indicated by the semi-transparent rectangle. The measurements
are taken from Ref. \cite{tegmark1}.}
\end{figure}
\end{center}

\begin{center}
\begin{figure}
\includegraphics[width=12cm]{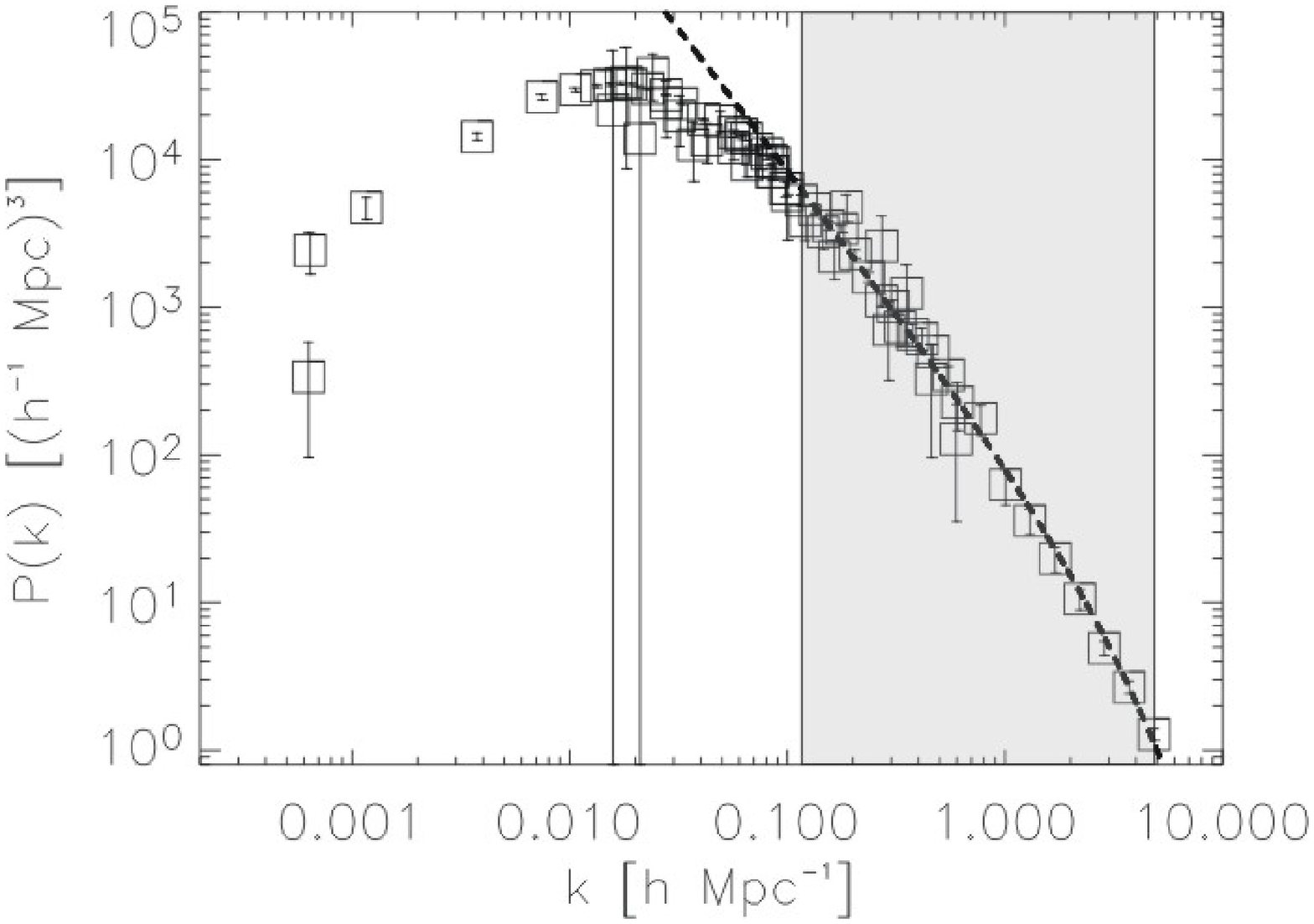}
\caption{\label{fig:cosmo_largek} Wavenumber spectrum of the
combined cosmological measurements. Squares are measured points.
The dashed line is a fit to Eq. (\ref{eq:exp_pow_decay}) using the
data indicated by the semi-transparent rectangle. The measurements
are taken from Ref. \cite{tegmark1}.}
\end{figure}
\end{center}

During our analysis of the measurements it became obvious that Eq.
(\ref{eq:exp_pow_decay}) does not fit the entire range of
wavenumbers above the peak of the spectrum. Therefore two fits
were made:

\begin{itemize}
\item A small wavenumber fit where the smallest wavenumber used is
chosen to be the one closest to (but larger than) the peak of the
$\Lambda$CDM model shown in Fig. \ref{fig:cosmo_allk}. The largest
wavenumber used is chosen so that $\chi^2$ of the fit is 1. In
Fig. \ref{fig:cosmo_smallk} the fit is displayed as a dashed line.
The parameters found are $n$ = 13.3 $\pm$ 1.6 h$^{-1}$ Mpc and
$\alpha$ = 0.24 $\pm$ 0.06. Note the interesting proximity of this
$\alpha$ to the result in section \ref{subsec:pci}.
\item A large wavenumber fit where the largest wavenumber used is
the maximum of the data set and the smallest wavenumber is chosen
so that $\chi^2$ of the fit is 1. In Fig. \ref{fig:cosmo_largek}
the fit is displayed as a dashed line. The parameters found are
$n$ = 0.33 $\pm$ 0.05 h$^{-1}$ Mpc and $\alpha$ = 1.90 $\pm$ 0.05.
\end{itemize}

The wavenumbers not being used in either fit define a transitional
region: [0.08, 0.10] h Mpc$^{-1}$. This interval is in rough
agreement with the transition found between the two power-law fits
shown in Fig. \ref{fig:sdss}.

\section{Revised analysis of stellarator fusion plasmas}
\label{sec:anal}

\begin{center}
\begin{figure}
\includegraphics[width=12cm]{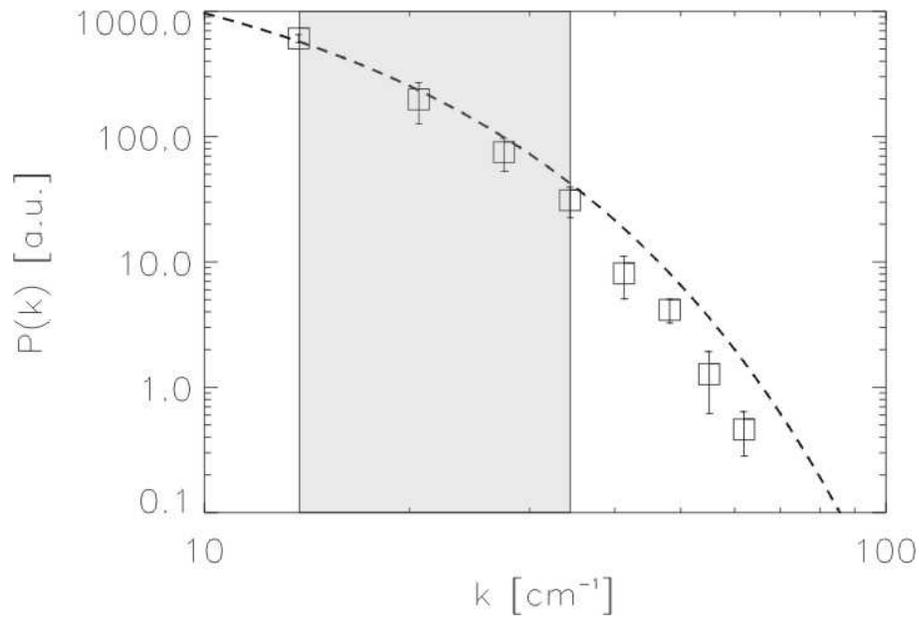}
\caption{\label{fig:lotus_lowk} Wavenumber spectrum of broadband
turbulence in W7-AS. Squares are measured points. The dashed line
is a fit to Eq. (\ref{eq:exp_pow_decay}) using the measurements
indicated by the semi-transparent rectangle. Note that this is
exclusively an amplitude fit, we use $n$ and $\alpha$ from the PCI
fit, see section \ref{subsec:pci} and Fig. \ref{fig:pci}.}
\end{figure}
\end{center}

\begin{center}
\begin{figure}
\includegraphics[width=12cm]{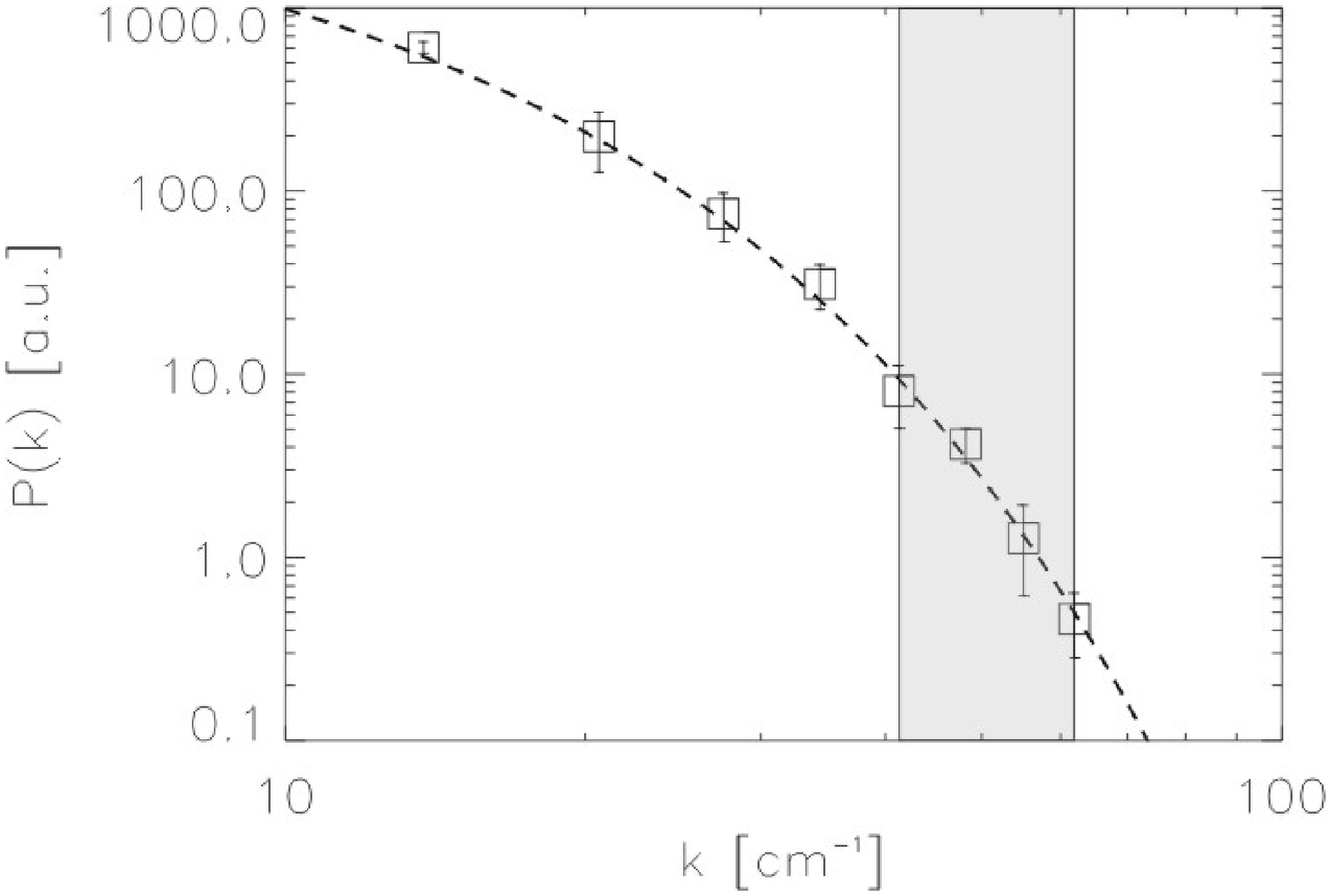}
\caption{\label{fig:lotus_highk} Wavenumber spectrum of broadband
turbulence in W7-AS. Squares are measured points. The dashed line
is a fit to Eq. (\ref{eq:exp_pow_decay}) using the measurements
indicated by the semi-transparent rectangle.}
\end{figure}
\end{center}

Based on the insights gained in section \ref{sec:add}, we revisit
the turbulence measurements made in W7-AS, see Fig.
\ref{fig:w7as}. Our approach is to make three fits, one for small,
one for large and one for all wavenumbers:

\begin{itemize}
\item For the fit to small wavenumbers we use $n$ and $\alpha$
found from the PCI measurements in section \ref{subsec:pci} and
only allow the overall amplitude to vary. The largest wavenumber
to be included in the fit is chosen so that $\chi^2$ of the fit is
1. Using this criterion the four smallest wavenumbers are
utilized. In Fig. \ref{fig:lotus_lowk} the fit is shown as a
dashed line.
\item We allow all fit parameters in Eq. (\ref{eq:exp_pow_decay})
to vary for the fit to the four largest wavenumbers and show the
resulting fit as a dashed line in Fig. \ref{fig:lotus_highk}. The
fit yields $n$ = 0.14 $\pm$ 0.008 cm and $\alpha$ = 0.25 $\pm$
0.1. This fit is also good for the entire wavenumber range: Using
the found parameters, $\chi^2$ remains less than 1 for all eight
wavenumbers.
\item Fit to Eq. (\ref{eq:exp_pow_decay}) using all eight
wavenumbers and allowing every fit parameter to vary. The fit
yields $n$ = 0.14 $\pm$ 0.007 cm and $\alpha$ = 0.11 $\pm$ 0.2,
which is consistent with a pure exponential function (Eq.
(\ref{eq:exp_decay_simpl})).
\end{itemize}

The amplitude-only fit using PCI parameters indicates a transition
at the center of the data set, whereas the two other fits suggest
that the data is consistent with a continuous behavior. It is
apparent that additional measurements at large wavenumbers in
fusion plasmas are needed to gain confidence in the fits. Some of
the existing points might be in a transitional region akin to the
one found in the cosmological data.

\section{Discussion}
\label{sec:disc}

The fact that density fluctuations on small (fusion plasmas) and
large (cosmological) scales can be described by an exponential
function with an algebraic pre-factor, Eq.
(\ref{eq:exp_pow_decay}), might indicate that plasma turbulence at
early times has been expanded to cosmological proportions. A
natural consequence of that thought would be to investigate
fluctuations in quark-gluon plasmas (QGPs) corresponding to even
earlier times. However, experimental techniques to do this are not
sufficiently developed at the moment due to the extreme nature of
QGPs. It has been suggested that complex (or dusty) plasmas could
be used as a model of QGPs \cite{thoma}.

It is fascinating that wavenumber spectra over wider scales peak
at small wavenumbers and decrease both above and below that peak.
This is seen both in fusion plasmas and on cosmological scales,
compare Figs. \ref{fig:pci} and \ref{fig:cosmo_allk}. Turbulence
theory in 1D or 3D would interpret the peak position as the scale
where energy is injected.

Fitting wavenumber spectra to power-laws as we did in section
\ref{sec:rev} is based on fluid turbulence theories, but in
general care must be taken when interpreting the outcome: We know
that an exponential function can be Taylor expanded to a Maclaurin
series

\begin{equation}
P(k) \propto e^{-nk} = \sum_{i=0}^{\infty} \frac{(-nk)^i}{i!}.
\label{eq:taylor}
\end{equation}

So locally, i.e. for a small range of wavenumbers, an exponential
dependency can be masked as a power-law; the exponent would vary
as a function of the wavenumber range selected. To some extent
this objection also holds true for our results using Eq.
(\ref{eq:exp_pow_decay}) to identify two wavenumber intervals
having different fit parameters. It is an open question whether an
extension of the cosmological data to larger wavenumbers would
necessitate further intervals. However, in fluid turbulence
simulations a transition between near- and far-dissipation similar
to the one we discuss has been identified \cite{martinez}. This
observation lends support to the interpretation that our dual fits
describe different types of dissipation.

To sum up, we favor Eq. (\ref{eq:exp_pow_decay}) as a descriptor
of the data, both for fusion plasma and cosmological measurements.
Perhaps forcing occurs at a large scale where the spectra peak and
transitions directly to dissipation. The fact that we obtain $n
\simeq \rho_s$ using Eq. (\ref{eq:exp_pow_decay}) for fusion
plasmas suggests that $n$ is the characteristic scale of
turbulence.

We would like to point out that observations of electric field
fluctuations in the solar wind are found to be proportional to
exp($-k \rho_i$/12.5) for $k \rho_i$ $>$ 2.5 \cite{bale1}. It
should be noted that the electric field measurements at these
large wavenumbers are noisy. Work on the solar wind data to
analyze density fluctuations is in progress \cite{bale2}.

\section{Conclusions}
\label{sec:conc}

We have in this paper reported on suggestive similarities between
density fluctuation power versus wavenumber on small (mm) and
large (Mpc) scales.

The small scale measurements were made in fusion plasmas and
compared to predictions from turbulence theory. The data sets fit
Eq. (\ref{eq:exp_pow_decay}), which has a functional form that can
be explained as dissipation by turbulence theory. The large scale
cosmological measurements can also be described by Eq.
(\ref{eq:exp_pow_decay}). In general, two wavenumber ranges
separated by a transitional region are identified.

The similar dependency of density fluctuation power on wavenumber
might indicate a common origin of these fluctuations, perhaps from
fluctuations in QGPs at early stages in the formation of the
universe. The value of $\alpha$ is almost identical for both
fusion plasma and cosmological measurements at wavenumbers close
to but above the peak of the spectra.

To progress further, it is essential that the quantity of
wavenumber-resolved fusion plasma turbulence measurements is
vastly increased.

\ack This work was supported at MIT by the Department of Energy,
Cooperative Grant No. DE-FC02-99ER54512. We thank M Tegmark for
providing all cosmological measurements analyzed in this paper.

\newpage

\clearpage

\clearpage

\clearpage

\clearpage

\clearpage

\clearpage

\clearpage

\end{document}